\documentclass[a4paper,11pt]{article}
\pdfoutput=1 %
\usepackage{jcappub} 

\usepackage[T1]{fontenc} 

\usepackage[utf8]{inputenc}

\usepackage{hyperref} 
\usepackage{url}
\usepackage{graphicx}
\usepackage{esvect}
\usepackage{graphics}
\usepackage{natbib}
\usepackage{mathrsfs}
\usepackage{blindtext}
\usepackage{comment} 
\usepackage{placeins} 

\usepackage[table]{xcolor}
\usepackage{multirow}

\usepackage[normalem]{ulem}
\usepackage{color}

\usepackage{soul}

\usepackage{supertabular}
\usepackage{tabularx}

\def\units#1{~\hbox{$\,{\rm #1}$}}

\title{Combined analyses of the antiproton production from cosmic-ray interactions and its possible dark matter origin}

\author{Pedro~De~La~Torre~Luque}
\emailAdd{pedro.delatorreluque@fysik.su.se}

\affiliation{The Oskar Klein Centre, Department of Physics, Stockholm University, AlbaNova\\
  SE-10691 Stockholm, Sweden}

\date{\today}

\abstract{Recent cosmic-ray (CR) studies have claimed the possibility of an excess on the antiproton flux over the predicted models at around $10\units{GeV}$, which can be the signature of dark matter annihilating into hadronic final states that subsequently form antiprotons.  However, this excess is subject to many uncertainties related to the evaluation of the antiproton spectrum produced from spallation interactions of CRs. In this work, we implement a combined Markov-Chain Monte Carlo analysis of the secondary ratios of B, Be and Li and the antiproton-to-proton ratio ($\Bar{p}/p$), while also including nuisance parameters to consider the uncertainties related to the spallation cross sections.   This study allows us to constrain the Galactic halo height and the rest of propagation parameters, evaluate the impact of cross sections uncertainties in the determination of the antiproton spectrum and test the origin of the excess of antiprotons. In this way, we provide a set of propagation parameters and scale factors for renormalizing the cross sections parametrizations that allow us to reproduce all the ratios of B, Be, Li and $\bar{p}$ simultaneously.

We show that the energy dependence of the $\Bar{p}/p$ ratio is compatible with a pure secondary origin. 
 Specifcally, we find that the energy dependence of the evaluated $\Bar{p}/p$ spectrum matches the AMS-02 data at energies above $\sim 3\units{GeV}$, although there is still a nearly constant $\sim10\%$ excess of $\Bar{p}$ over our prediction. We discuss that this discrepancy is more likely explained from a $\sim10\%$ scaling in the cross sections of antiproton production, rather than a component of dark matter leading to antiprotons. In particular, we find that the best-fit WIMP mass ($\sim 300 \units{GeV}$) needed to explain the discrepancy lies above the constraints from most indirect searches of dark matter and the resultant fit is poorer than with a cross sections scaling.
 
}

\begin{document}
\maketitle
\flushbottom

\section{Introduction}
\label{sec:intro}

Cosmic rays (CRs) constitute a valuable tool for studying the galactic environment (properties of interstellar plasma~\cite{Fornieri:2020wrr}, magnetic field configurations~\cite{Electrons_Bfield, B_CRs}, astrophysical sources and their surroundings~\cite{Sudoh:2019lav, Gamma_sou}, etc). After being injected from astrophysical sources (mostly supernova remnants~\cite{bell1978acceleration, axford1982structure,blandford1978particle}) CRs propagate in the Milky Way and produce other nuclei (called secondary CRs) when they hit gas in the interstellar medium (ISM). While secondary CRs are used to characterize the diffusion process that CRs experience along their way toward the Earth (due to collision-less interactions with plasma waves), other secondary particles like gamma-rays and neutrinos provide useful information about the sources that accelerate CRs, since the trajectories of these particles are not affected by magnetic fields and plasma waves. 
 
On top of this, CR antinuclei and positrons represent an useful tool for revealing the presence of the evasive dark matter (DM), since it is theoretically expected that they annihilate or decay into standard model particle-antiparticle pairs. Particularly interesting candidates are WIMPs~\cite{BERTONE2005279}, which have a mass sufficient to be able to produce nuclei-antinuclei pairs. In fact, many previous studies have tried to explain unexpected discrepancies between the predicted and measured spectra of positrons and antiprotons by including the expected production of these antiparticles from WIMPs. Nevertheless, there are yet large uncertainties related to the production of these antiparticles that could explain these discrepancies without any need of adding a dark matter component. The best example of this fact is the high-energy positron ``excess'', which was once considered evidence of dark matter annihilation, but is now more commonly considered to be a product of primary $e^+ e^-$ acceleration from pulsars~\cite{Hooper:2008kg, Manconi:2020ipm, SERPICO20122}.


Several studies have recently claimed an excess of antiprotons over the expected flux at $10-20 \units{GeV}$, assuming they are only originated as secondary particles, and have explained it by adding annihilation or decay from a WIMP~\cite{Reinert:2017aga, Cui:2018klo, Cuoco, Heisig:2020jvs, Cholis}. This excess seems to be compatible with a $50-100 \units{GeV}$ WIMP with a thermally-averaged cross section near the thermal cross section ($\left<\sigma v \right> \sim 3\times10^{-26} \units{cm^{3}/s}$) annihilating to hadronic final states ($q\bar{q}$ pairs, and mainly to $b\bar{b}$). Interestingly, many authors claimed that this WIMP seems to be compatible with the properties of the Galactic Center Excess (GEC) of gamma-rays observed by Fermi-LAT~\cite{DiMauro:2021raz, DiMauro:2021qcf} and has no tension with other direct or indirect DM searches~\cite{Hooper:2019xss, Leane:2020liq}. Nevertheless, a recent study shows that this signals can not be compatible with the most recent measurements~\cite{DiMauro:2021qcf}.

The total source term for secondary production of antiprotons from spallation reactions of CRs with the interstellar gas has the following form~\cite{Kappl:2014hha}:
\begin{equation}
Q_{CR+ISM\longrightarrow\bar{p}}(E_{\bar{p}}) = \sum_{CR}4 \pi~ n_{ISM} \int^{\infty}_0 dE \phi_{CR}(E)\frac{d\sigma_{CR+ISM\longrightarrow\bar{p}}}{dE_{\bar{p}}}(E, E_{\bar{p}}) ,
\label{eq:sec_AP_Sourceterm}
\end{equation}
where the key ingredients are the CR fluxes (essentially H and He) and the $\bar{p}$ production cross sections. In particular, the determination of the cross sections of $\bar{p}$ production is subject to important uncertainties, and results in $12-20\%$ uncertainties in the evaluation of the local $\bar{p}$ spectrum~\cite{Korsmeier, Cuoco2, Di_Mauro_Ap}. 

Further uncertainties in the evaluation of the antiproton flux come from the determination of the propagation parameters (essentially, the diffusion coefficient used to characterize the diffusive movement of CRs and the effective Alfv\'en velocity). Besides, the uncertainties related to the cross sections for secondary CR production are high ($> 20\%$), which leads to large uncertainties in the determination of the diffusion coefficient~\cite{Luque:2021joz, Luque:2021nxb, ICPPA_Pedro}, since this is usually determined from the flux ratios of secondary-to-primary CRs. Also, correlated errors in the AMS-02 data seem to lead to looser constraints in the determination of the propagation parameters, reducing the significance of the antiproton excess even below $1\sigma$ level and implying that current models are consistent with the AMS-02 antiproton data~\cite{Heisig, Boudaud}. The problem is that the correlation matrix is not yet public and the only way to proceed is to guess its terms. 
 Additionally, the uncertainties associated to solar modulation can affect the significance of this signal~\cite{Ap_Lin}. Then, once antiprotons are produced, they also interact with the interstellar gas as they propagate through the Galaxy, resulting either in annihilation or loss of a fraction of their energy. Non-annihilation inelastic interactions of antiprotons with interstellar protons yield an extra source of lower energy antiprotons, which we refer to as a ``tertiary'' source of antiprotons, but this represents a very subdominant contribution to the total spectrum of antiprotons~\cite{DRAGON2-2} so that the uncertainties related to this tertiary antiprotons is negligible for the evaluation of the antiproton spectrum.

In addition to the secondary antiproton production, dark matter can also produce a significant antiproton flux. the source term of antiprotons from DM annihilation (or decay) mainly depends on the DM density profile in the Galaxy $\rho (\vec{x})$, on the velocity-averaged DM annihilation cross section (or on the decay rate) into pairs of Standard Model particles ($q\bar{q}$ pairs, $W^+W^-$, etc.), and on the DM particle mass, as shown in the following equation~\cite{Fornengo:2013xda}:
\begin{equation}
Q^{Ann}_{DM\longrightarrow\bar{p}}(\vec{x}, E) = \frac{1}{2}\left(\frac{\rho (\vec{x})}{m_{DM}}\right)^2 \sum_f \left< \sigma v \right>_f \frac{dN_{\bar{p}}^f}{dE}
\label{eq:DM_AP_Sourceterm}
\end{equation}

In this paper, we investigate the antiproton spectrum produced from CR collisions with the interstellar gas by employing an analysis that combines the $\bar{p}/p$ spectrum with the flux ratios of secondary CRs B, Be and Li to constrain the propagation parameters and accounting for systematic uncertainties in the cross sections parametrizations of production of these secondary nuclei. This combined analysis allows us to reduce the grammage needed to reproduce the spectra of B, Be and Li, therefore enhancing the production of antiprotons from CR collisions with gas. In this way, we examine the $\Bar{p}/p$ ratio in the context of a combined analysis that also matches other ratios of secondary CRs.
We find that our prediction greatly reproduces the energy dependence of the $\bar{p}$ spectrum reported by AMS-02 above $\sim3\units{GeV}$ albeit with a constant $10\%$ offset compared to the data. We find that the use of the new AMS-02 $\bar{p}/p$ data is very relevant here and conclude that the $\bar{p}/p$ spectrum is compatible with a pure secondary origin of antiprotons. On top of this, we discuss that the origin of this $10\%$ discrepancy can be more plausibly explained by a scaling of the $\bar{p}$ cross sections, which offers a better fit to data with respect to a WIMP signal producing antiprotons.

\section{Methodology}
\label{sec:Method}

\subsection{Simulations setup}
\label{sec:Simsetup}

The propagation parameters have been usually inferred from the B/C spectrum~\cite{maurin2001cosmic}, nevertheless, nowadays AMS-02 data also provides unprecedented precision for the measurements of the flux ratios of the secondary CRs Be and Li~\cite{AMS_gen, Aguilar:2015ooa, Aguilar:2018keu, Aguilar:2020ohx, aguilar2017observation, aguilar2018observation, aguilar2019towards, Aguilar:2021tla}. The difficulty in obtaining a precise evaluation of these parameters, instead, resides in the large uncertainties related to the cross sections for the production of secondary CRs. These cross sections are incorporated in CR propagation codes via parametrizations of the many reaction channels involved in the CR nuclear network~\cite{Tomassetti:2017hbe}.
Many recent works have studied the impact of these uncertainties and stressed their importance for correctly evaluating the spectra of secondary CRs~\cite{Luque:2021joz, GenoliniRanking, Weinrich:2020cmw, Weinrich_halo, Korsmeier:2021brc}. A combination of different CR observables would allow us to reduce the effect of uncertainties in the determination of the propagation parameters~\cite{maurin2010systematic}. 

In this work, we implement a cylindrically symmetric (2-dimensional) model of the Galaxy. The general formulae describing CR propagation in the Galaxy are given by:
\begin{equation}
\label{eq:caprate}
\begin{split}
 \vec{\nabla}\cdot(\vec{J}_i - \vec{v}_{\omega}N_i) + \frac{\partial}{\partial p} \left[p^2D_{pp}\frac{\partial}{\partial p} \left( \frac{N_i}{p^2}\right) \right] = Q_i + \frac{\partial}{\partial p}  \left[\dot{p} N_i - \frac{p}{3}\left( \vec{\nabla}\cdot \vec{v}_{\omega} N_i \right)\right] \\
  - \frac{N_i}{\tau^f_i} + \sum \Gamma^s_{j\rightarrow i}(N_j) - \frac{N_i}{\tau^r_i} + \sum \frac{N_j}{\tau^r_{j\rightarrow i}} ,
\end{split}
\end{equation}
which takes into account CR diffusion in space (characterized by the diffusion coefficient, D) and in momentum (also known as reacceleration, characterized by $D_{pp}$), convection or advection, energy losses, injection from sources, decays and collisions with the ISM. These equations are numerically solved with the recent DRAGON2 code~\footnote{\label{note1} Available at \url{https://github.com/cosmicrays/DRAGON2-Beta\_version}}. More information about the code and the propagation equation can be found on refs.~\cite{DRAGON2-1} and~\cite{DRAGON2-2}.
The diffusion equation employed in this work has a rigidity-dependence modeled as:
\begin{equation}
 D (R) = D_0 \beta^{\eta}\frac{\left(R/R_0 \right)^{\delta}}{\left[1 + \left(R/R_b\right)^{\Delta \delta / s}\right]^s} ,
\label{eq:diff_eq}
\end{equation}
where the reference rigidity is set to $R_0 = 4\units{GV}$ and the values of the rigidity break ($R_b$), the change in spectral index ($\Delta\delta = \delta - \delta_h $, where $\delta_h$ is the spectral index for rigidities above the break), and the smoothing parameter $s$ are: $\Delta\delta = 0.14 \pm 0.03$, $R_b = (312 \pm 31) \units{GV}$ and $s = 0.040 \pm 0.0015$, as determined in Ref~\cite{genolini2017indications} from a detailed analysis of the proton and helium AMS-02 fluxes.

The spatial diffusion coefficient and diffusion coefficient in momentum space are related by the Alfv\'en velocity, $V_A$~\cite{osborne1987cosmic, seo1994stochastic}:
\begin{equation}
    D_{pp} (R) = \frac{4}{3}\frac{1}{\delta (4 - \delta^2)(4 - \delta)} \frac{V_A^2 p^2}{D(R)} . 
    \label{dpp} 
\end{equation}

To simulate the solar modulation effect on CRs reaching the Earth, we make use of the Force-Field approximation~\cite{forcefield}, characterized by the Fisk potential, $\phi$. Charge-sign modulation is applied here, following the approach developed in Ref~\cite{Cholis:2015gna}. In this case, the solar modulation potential is modified to have the form:
\begin{equation}
\phi^{\pm} (t, R) = \phi_0(t) + \phi_1^{\pm}(t) F(R/R_0) ,
\label{eq:Charge-sign_Modul}
\end{equation}
choosing $F(R/R_0) \equiv \frac{R_0}{R}$ and $R_0 = 1 \units{GV}$, similarly to what was done in Ref~\cite{Reinert:2017aga}. We set $\phi^+_{1,AMS-02} = 0 $ and explore values of $\phi^-_{1}$ around $0.9 \units{GV}$, close to that derived in~\cite{Reinert:2017aga}. $\phi_0$ is the usual Fisk potential, which we set to be $0.61 \units{GV}$, since it allows us to reconcile the Voyager-1~\cite{stone2013voyager,cummings2016galactic} and AMS-02 data in the period 2011-2016 (period of data collection for the AMS-02 data on secondary-to-primary ratios, He, C and O). This value is also chosen to be consistent with the data from the NEWK neutron monitor experiment\footnote{\url{http://www01.nmdb.eu/station/newk/}} (see~\cite{ghelfi2016non,ghelfi2017neutron}). The AMS-02 proton data was collected in the period of 2011-2013, leading to a difference in the Fisk potential $\sim 0.1 \units{GV}$, so that to compute the proton flux at Earth we use $\phi_0 = 0.60 \units{GV}$. The new AMS-02 data on the $\bar{p}/p$ spectrum~\cite{AGUILAR20211} is obtained from the period 2011-2018, so that we employ a value of $\phi_0 = 0.58 \units{GV}$ for the evaluation of its spectrum ($\Delta\phi$ is calculated using NEWK data from both periods, following~\cite{ghelfi2016non,ghelfi2017neutron}).

\subsection{Analysis setup}
\label{sec:Ansetup}
In this work, we perform combined analyses of the secondary CRs: B, Be and Li, combined with antiprotons ($\bar{p}$), in order to study the predicted $\bar{p}/p$ spectrum from CR interactions with the ISM gas and to study a possible signal of a WIMP-produced $\bar{p}$ in the Galaxy. The propagation parameters (H, $D_0$, $V_A$, $\eta$ and $\delta$) are inferred from the flux ratios of these CR species by means of a Markov chain Monte Carlo (MCMC) procedure, that relies on Bayesian inference.
This analysis determines the probability distribution functions for the propagation parameters studied in this work to describe the AMS-02 data and their confidence intervals. This algorithm was presented in~\cite{Luque:2021nxb}, but it has been upgraded to be able to include the halo height in the procedure. The analysis also incorporates nuisance parameters (scale factors) to further adjust the normalization of the spallation cross sections of B, Be and Li production ($\mathcal{S}_{B}$, $\mathcal{S}_{Be}$, $\mathcal{S}_{Li}$), which is the main source of uncertainty in the evaluation of secondary-to-primary flux ratios. The injection parameters of primary CRs $^1$H, $^{4}$He, $^{12}$C, $^{14}$N, $^{16}$O, $^{20}$Ne, $^{24}$Mg and $^{28}$Si are adjusted to reproduce the AMS-02 data.  These parameters are readjusted after a set of propagation parameters is obtained from the MCMC analysis, and we repeat again the analysis. We iterate in this way until the algorithm converges. This is mainly relevant for the evaluation of the $\bar{p}/p$ spectrum, for which a good adjustment of the proton and He spectra are important. For more details on the MCMC analysis, we refer the reader to Ref~\cite{Luque:2021nxb}. We take into account the production of antiprotons from nuclei heavier than He (namely, C, N, O, Ne, Mg and Si) by scaling the proton cross sections (both, cross sections of interactions with p and He as targets in the ISM) by a $A^{2/3}$ factor, where $A$ is the mass number of the CR nuclei colliding with ISM gas. The contribution from these nuclei implies a $\sim 5-6\%$ of the total antiproton flux between $10 \units{GeV/n}$ and $100 \units{GeV/n}$, in agreement with what was found in Ref.~\cite{Korsmeier}. 
Furthermore, we stress that we employ here the new AMS-02 data for the $\bar{p}/p$ spectrum~\cite{AGUILAR20211}.

Recently, the authors of Ref~\cite{Luque:2021nxb} showed that the secondary-to-primary flux ratios of B and Be infer roughly identical values of the normalization ($D_0$) and spectral index ($\delta$) of the diffusion coefficient (eq.~\ref{eq:diff_eq}) using the {\tt DRAGON2} parametrization after rescaling their cross sections according to the AMS-02 Be/B, Li/B and Li/Be flux ratios. Nevertheless, the ratios of Li show significant differences in the determination of the propagation parameters with respect to B and Be ratios. In addition, it was demonstrated that this parametrization yield the best agreement to AMS-02 data in a combined analysis of B, Be and Li flux ratios. Therefore, the {\tt DRAGON2} parametrizations are used for the production of heavier secondary CRs~\cite{Luque:2021joz} and the analyses are focused on the secondary-to-primary ratios of B and Be. The cross sections derived in Ref~\cite{Winkler:2017xor} for the $\Bar{p}$ production cross sections are used for the evaluation of the antiproton spectrum.

Different analyses have been performed: The ``Standard analysis'', which includes the B/C, B/O, Be/C, Be/O flux ratios (which are able to constrain the values of $D_0$/H, $V_A$, $\eta$ and $\delta$), the Be/B, Li/B, Li/Be ratios (which allow setting the values of the scale parameters) and the $^{10}Be$/Be and $^{10}Be$/$^{9}Be$ flux ratios (allowing us to constrain the halo height, H~\cite{Donato:2001eq}). Then, we carry out another analysis including also the $\Bar{p}/p$ ratio (taking into account antiprotons only produced from CR interactions with the ISM gas, eq.~\ref{eq:sec_AP_Sourceterm}). Nevertheless, it must be mentioned that in these analyses we do not have into account correlations in the AMS-02 data, since the public data from the Collaboration do not include the full covariance matrix~\cite{Weinrich:2020cmw, Heisig}.


The combination of all these CR observables allow us to constrain the galactic halo height and the rest of propagation parameters, evaluate the impact of the spallation cross sections uncertainties in the determination of the antiproton spectrum and test the excess of antiprotons at $\sim 10 \units{GeV}$. Other works have also included the uncertainties related to the spallation cross sections, although they only used the B/C ratio and a refit to the very uncertain cross sections measurements of B production (with $1\sigma$ uncertainties $>20\%$, in average~\cite{Luque:2021joz}). 
Here, in turn, we make use of the B and Be CR species and the scale factors calculated by combining the high energy part of the flux ratios among B, Be and Li with the cross sections data.   Furthermore, these works fix the value of the halo height and do not include it in their analyses (mainly due to the huge uncertainties related to its determination). 

\subsubsection{Computation of a WIMP contribution}
\label{sec:Wimpsetup}
Finally, we also perform an analysis including the possible annihilation of a dark matter particle (a WIMP) into hadronic states that produce $\bar{p}$ (eq.~\ref{eq:DM_AP_Sourceterm}), in order to infer the WIMP mass and thermally averaged annihilation cross sections ($\left< \sigma v \right>$) needed to explain the discrepancies found assuming a pure secondary origin of antiprotons. This analysis is focused in the $\chi\,\rightarrow\, b\bar{b}$ annihilation channel.
Two different DM density profiles are analysed to assess the influence of this component in the determination of the WIMP mass and thermally averaged annihilation cross sections. Among the most used DM profile densities, derived from N-body simulations and cosmological studies~\cite{Ascasibar:2003mm, Lu:2005tu}, the most common one is the Navarro-Frenk-White (NFW) density profile~\cite{Navarro:1995iw}, which we use as reference. Other common density profiles are, the Einasto profile~\cite{Einasto, Graham:2005xx}, Isothermal profile~\cite{Isothermal}, Moore profile~\cite{Moore}, or the Burkert profile~\cite{Burkert:1995yz, Bur_Gal}. In order to compare with the NFW profile, we use the Burkert profile, since they present the largest differences, as can be seen from figure 1 of ref~\cite{Cirelli:2010xx}. These profiles are generally given by:
\begin{equation}
    \rho_{NFW}(r) = \rho_h \frac{r_h}{r} \left(1 + \frac{r}{r_h}\right)^{-2}
\label{eq:profiles_NFW}
\end{equation}
\begin{equation}   
    \rho_{BUR}(r) = \frac{\rho_h}{(1 + r/r_h)(1 + (r/r_h)^2)} , 
\label{eq:profiles_BUR}
\end{equation} 
with a characteristic halo radius $r_h\sim20 \units{kpc}$ , and a characteristic halo density $\rho_h$, used to normalize the profiles from observations~\cite{Salucci:2010qr}. In particular, we normalized these profiles to a value $\rho_{\odot} = 0.43\units{GeV}$ at the Solar System position, $r_{\odot} = 8.3\units{kpc}$.
We use DarkSusy~\cite{Bringmann:2018lay} to compute the antiproton production from WIMPs at Earth.

\section{Results}
\label{sec:results}

\subsection{Standard analysis}
\label{sec:Standard}

This analysis focuses on the combination of the ratios of secondary CRs mentioned above. The diffusion parameters obtained from this analysis are given in table~\ref{tab:diff_params}, along with their statistical uncertainties. As we see, the best fit halo height, H, is around $7\units{kpc}$, compatible with recent analyses~\cite{Luque:2021joz, Weinrich_halo, CarmeloBeB} and the ratio $D_0 \, (10^{28}\units{cm^{2}/s}) / H\, (kpc)$ is close to $1$. We must mention here that the determination of the $H$ value is very uncertain, since it is determined from the ratios of $^{10}$Be, for which most of experimental data are located around a few hundred MeV/n. This energy region is particularly affected by the uncertainties related to the solar modulation effect and the diffusion coefficient at such energies - the low energy part of the diffusion coefficient is the most uncertain, since this energy region is particularly complex, given the effect of damping and dissipation of plasma waves that seem to become important at sub-GeV energies~\cite{Reichherzer:2019dmb, Ptuskin_2006} and the fact that there are very few data points in the secondary-to-primary spectra below $1 \units{GeV/n}$. On top of this, at energies around a few hundreds of MeV/n the production of secondary CRs is dominated by nuclear resonances.
The value found for the effective Alfv\`en velocity, $V_A$, is $\sim 22 \units{km/s}$, in agreement with the expected values derived in Ref~\cite{Spangler:2010nu}. In addition, the value of the spectral index of the diffusion coefficient, $\delta$, is $\sim 0.43$, well compatible with other recent analyses~\cite{Boschini2020, Korsmeier:2021brc}. The determination of this parameter is also affected by an extra systematic uncertainty around $\pm 0.01$, since we fixed the value of the spectra index after the break (and break position), as discussed in~\cite{Luque:2021nxb}.

\begin{figure}[!t]
\centering
a)\includegraphics[width=0.49\textwidth,height=0.255\textheight,clip]{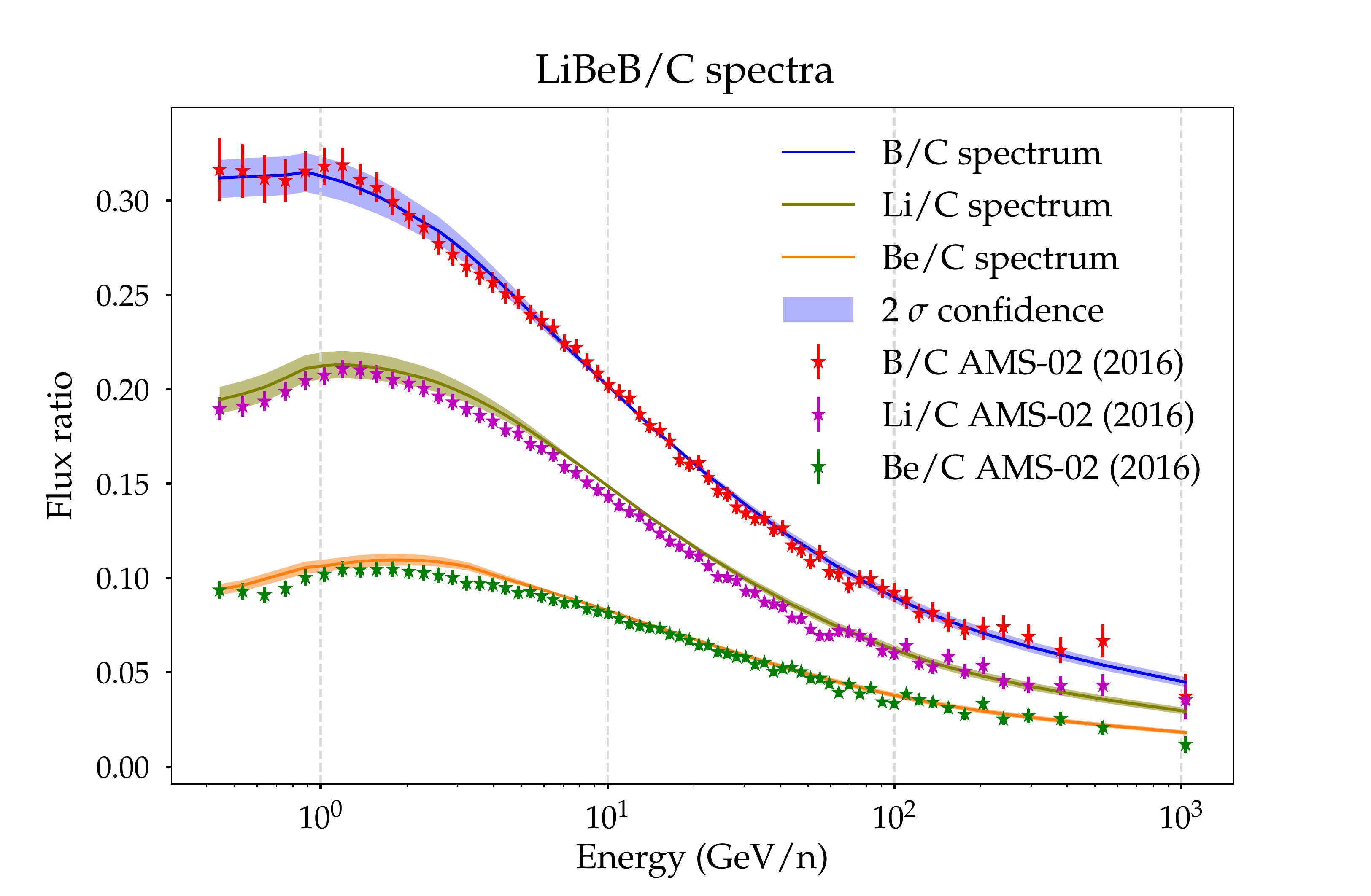} 
\hspace{-0.65cm} b)\includegraphics[width=0.49\textwidth,height=0.255\textheight,clip]{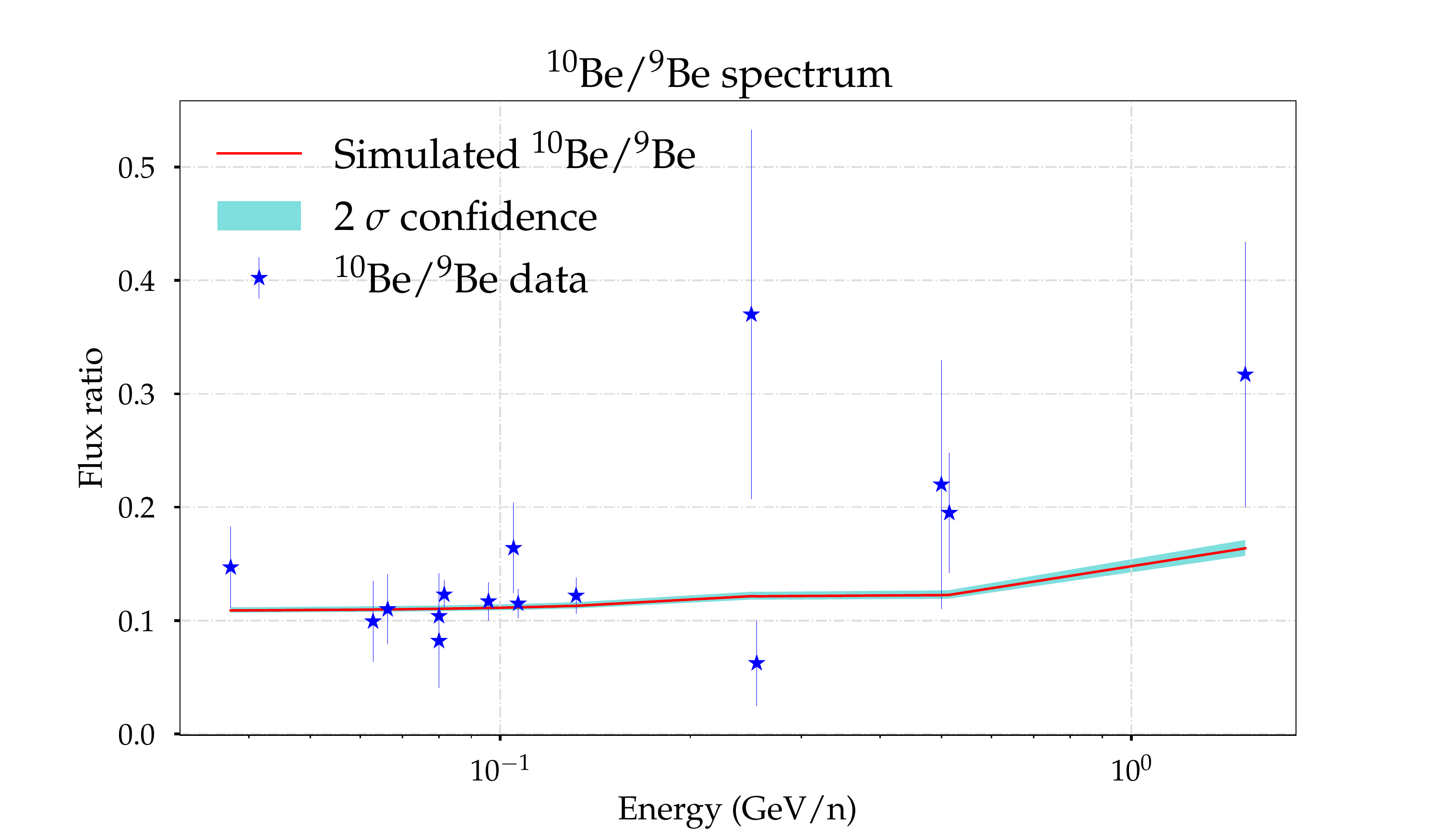}

c)\includegraphics[width=0.49\textwidth,height=0.25\textheight,clip]{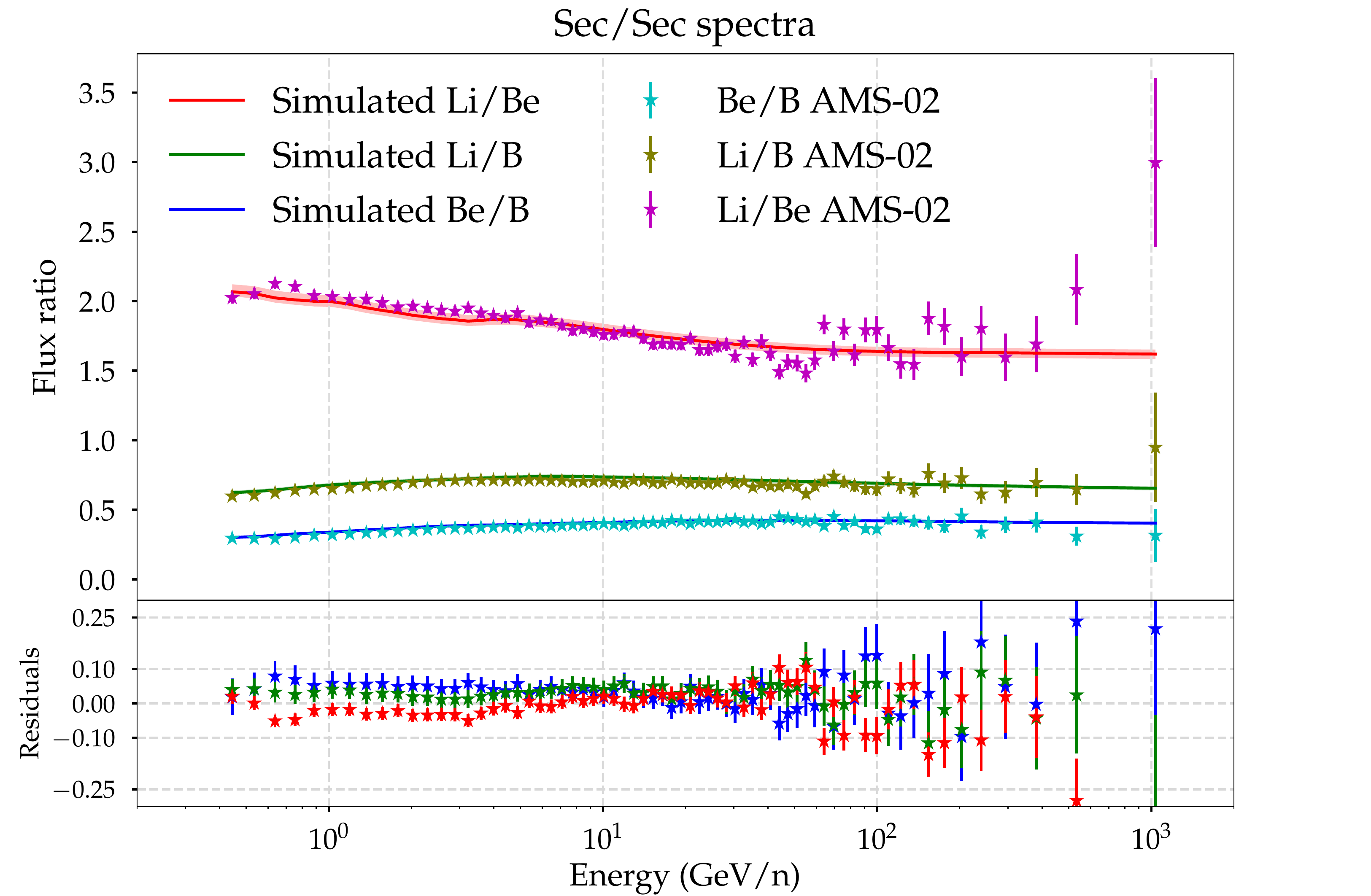}
\hspace{-0.6cm} d)\includegraphics[width=0.49\textwidth,height=0.25\textheight,clip]{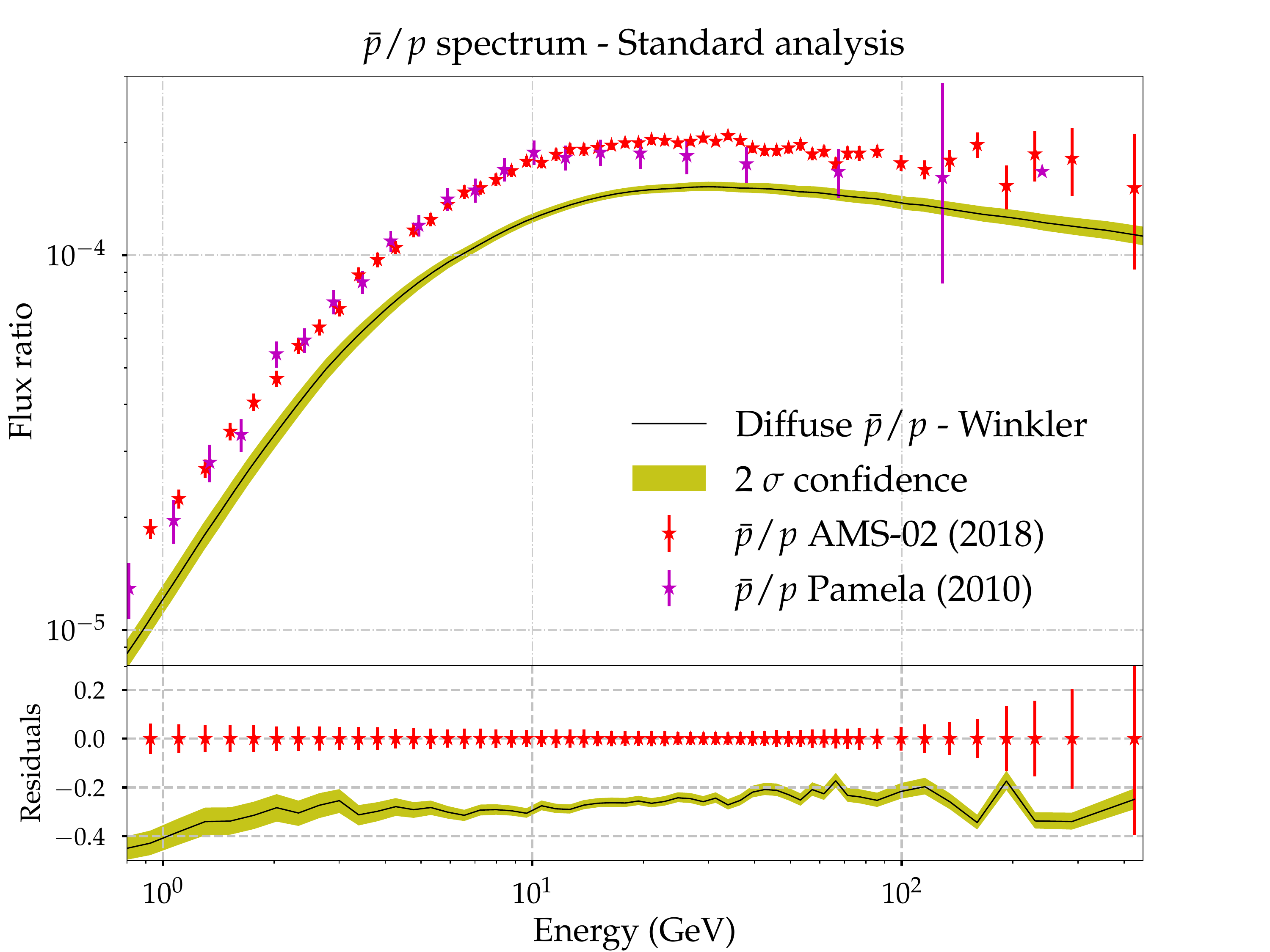} 
\caption{Relevant ratios involving secondary CRs produced in CR interactions with interstellar gas evaluated with the propagation parameters obtained in the {\it Standard analysis} and compared to experimental data. Panel a: flux ratios of B, Be and Li to C. Panel b: $^{10}$Be/$^9$Be flux ratio. Panel c: secondary-to-secondary flux ratios of B, Be and Li and the residuals with respect to AMS-02 data, defined as (model-data)/data. Panel d: $\bar{p}/p$ flux ratio evaluated with the Winkler cross sections and the residuals with respect to AMS-02 data. Error bands represent the $2\sigma$ statistical uncertainty.}
\label{fig:Standard}
\end{figure}

Interestingly, the scale factors derived in the analyses, as nuisance parameters, are very small: a $5-6\%$ scaling is found for the cross sections of B production and a $\sim 1\%$ scaling for Be and Li. We note that, while the related $1\sigma$ statistical uncertainty is of $\pm0.1$, the systematic uncertainty related to these scale factors has been estimated to be around $10\%$, with a $\sim5\%$ uncertainty associated to multi-step reactions and a $\sim3\%$ due to inelastic cross sections~\cite{Luque:2021nxb}. The gas distribution employed in the simulations can also add an additional $4\%$ uncertainty in the determination of these scale factors.

As we can see from panels a, b and c of figure~\ref{fig:Standard}, the ratios of the secondary CR nuclei can be fitted simultaneously within the AMS-02 uncertainty ($<5\%$). Nevertheless, the evaluation of the $\bar{p}/p$ with these propagation parameters yield a $20-30\%$ discrepancy with respect to the AMS-02 data, as shown in~\ref{fig:Standard}, d. 
Experimental data are taken from the ASI Cosmic Ray Data Base~\cite{Pizzolotto:2017lfd}~\footnote{\url{https://tools.ssdc.asi.it/CosmicRays/}} and the Cosmic-Rray DataBase~\cite{2020Univ....6..102M}~\footnote{\url{https://lpsc.in2p3.fr/crdb/}}.

\subsection{Combined analysis of the $\bar{p}/p$ spectrum}
\label{sec:Ap_analysis}

\begin{figure}[!t]
\centering
\includegraphics[width=0.52\textwidth,height=0.255\textheight,clip] {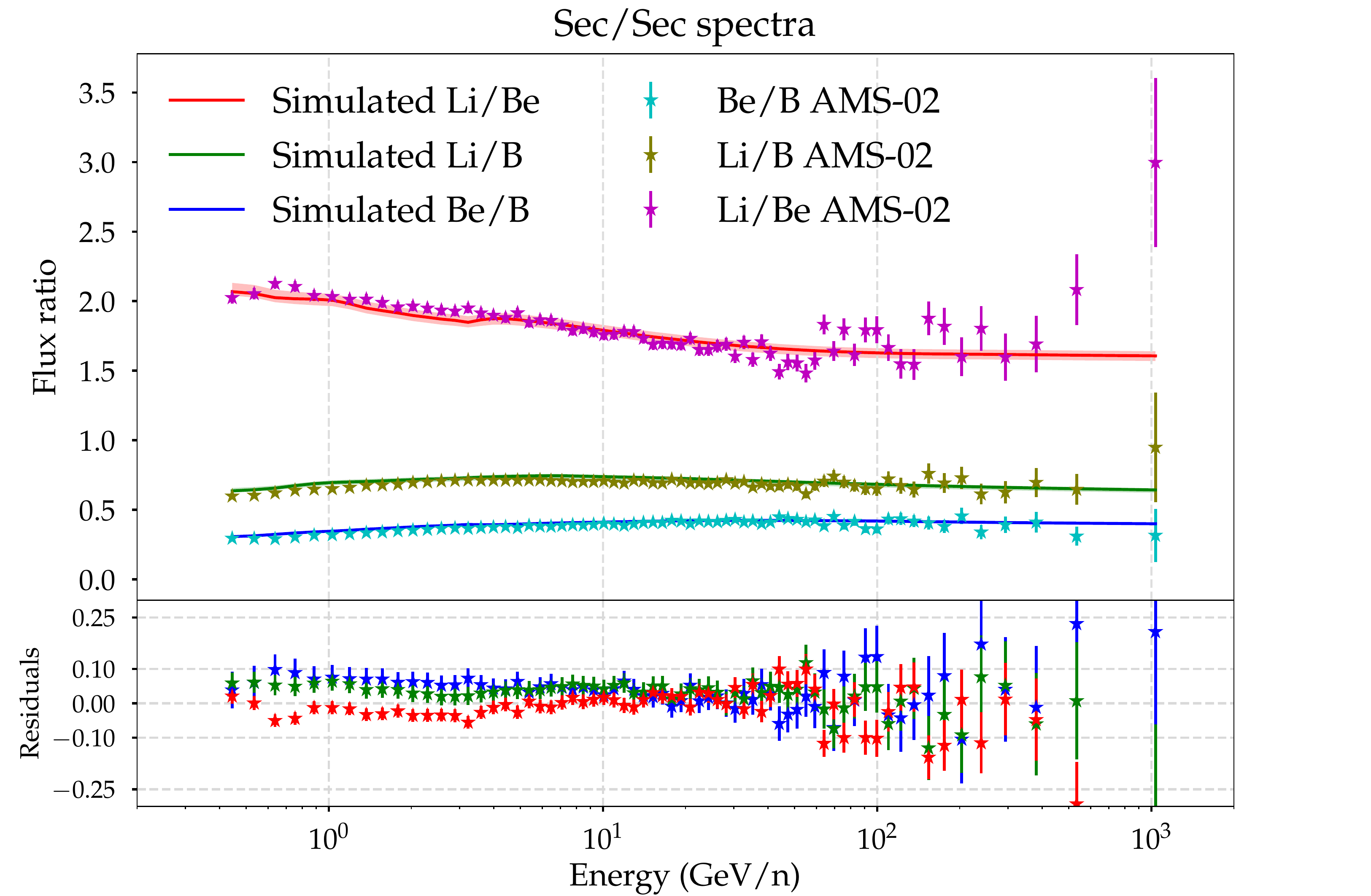} 

\includegraphics[width=0.513\textwidth,height=0.255\textheight,clip]{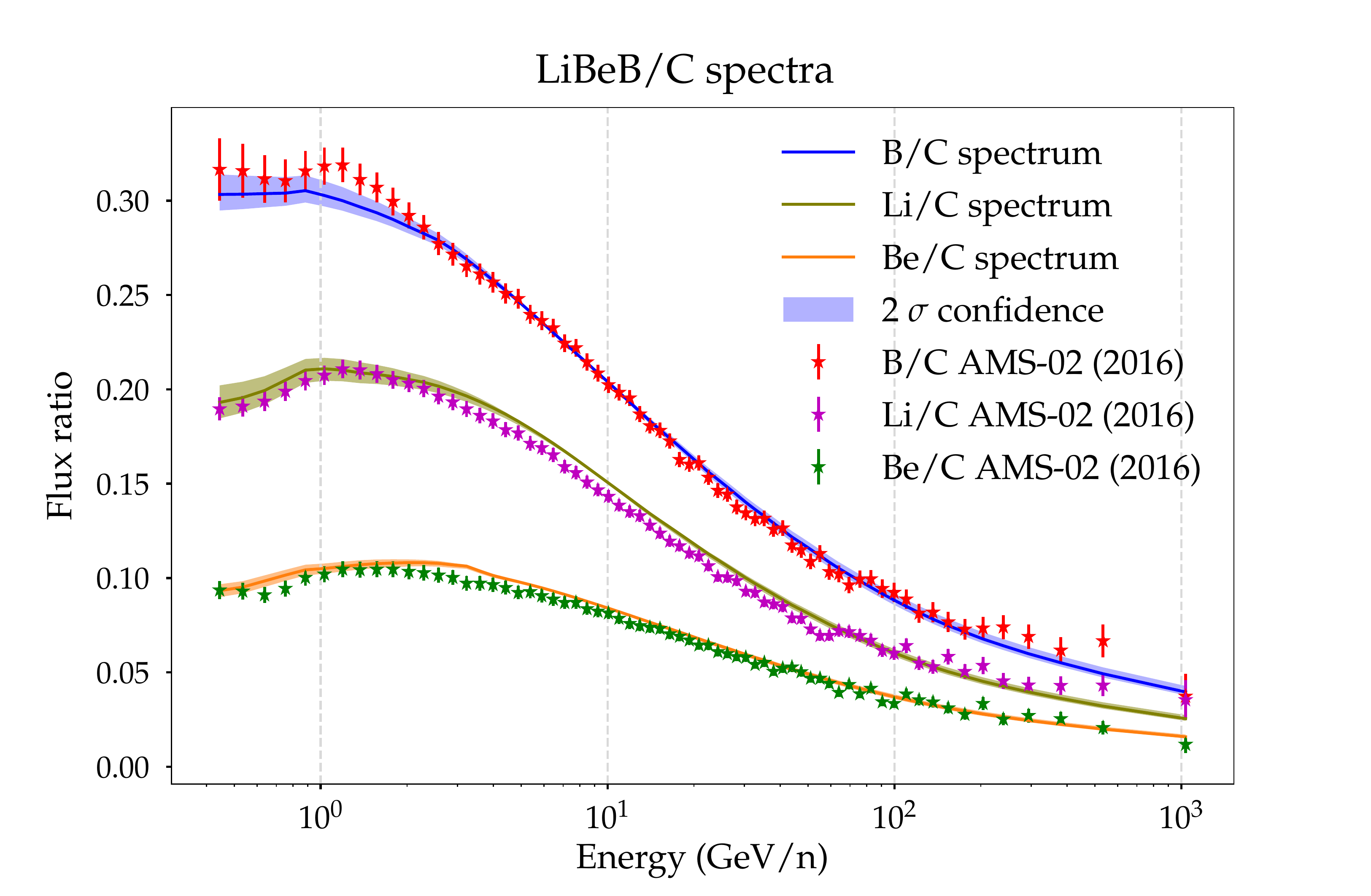} 
\hspace{-0.6cm}\includegraphics[width=0.513\textwidth,height=0.255\textheight,clip]{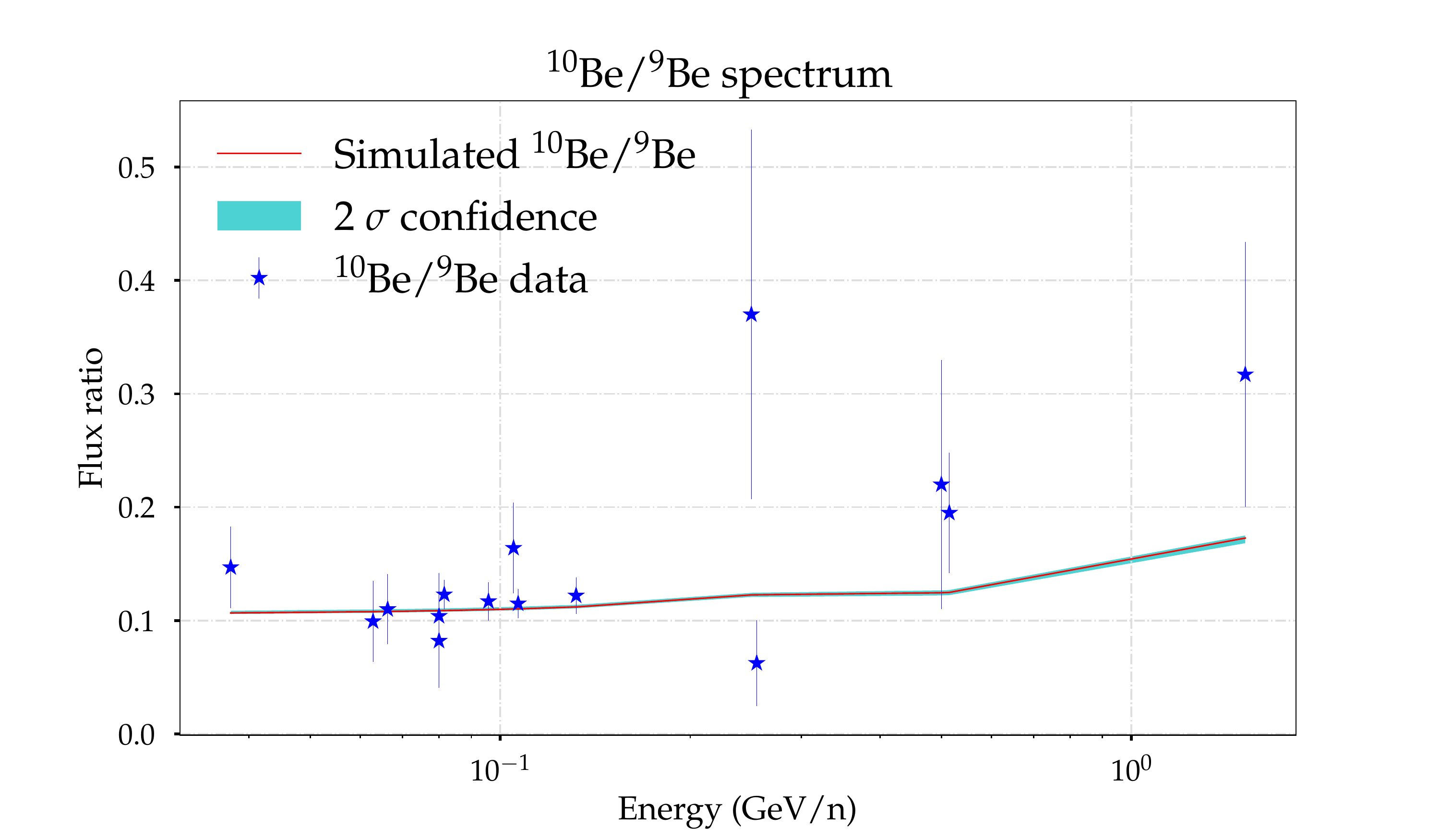}
\caption{ Similar to panels a, b and c of figure~\ref{fig:Standard} but for the propagation parameters determined in the combined analysis with the $\bar{p}/p$ ratio.}
\label{fig:Winkler_ratios}
\end{figure}

In this section, we report the results obtained from the combined analysis of the flux ratios included in section~\ref{sec:Standard}, adding also the $\bar{p}/p$ ratio above $4\units{GeV}$ in the analysis, without including additional nuisance parameters for the cross sections of production of $\bar{p}$ (i.e. not modifying the original cross sections). We have chosen $4\units{GeV}$ to reduce the impact of the solar modulation uncertainties in the evaluation of the $\bar{p}/p$ ratio and performed the analysis also with the $\bar{p}/p$ spectrum from energies from $3$ and $5\units{GeV}$, obtaining no significant differences in the evaluated propagation parameters above $1\sigma$. As we see from figure~\ref{fig:Winkler_ratios}, the ratios of secondary CR nuclei are still compatible with AMS-02 data within $2\sigma$ statistical uncertainty. The scale factors play a major role in this analysis, allowing us to readjust the grammage and diffusion parameters (mostly determined by the ratios B/C, B/O, Be/C and Be/O) to improve our predicted $\Bar{p}/p$ spectrum while keeping a good agreement to the other secondary ratios. In fact, these parameters change by up to $10\%$ with respect to the values predicted from the {\it Standard analysis} (see table~\ref{tab:diff_params}). We remind the reader that the uncertainties associated to the cross sections measurements are $> 20\%$ for the main channels of production of Be and B. We also observe that the value of $\delta$ inferred from this analysis is around $0.49$, closer to a Kraichnan spectrum for the turbulence. 
 In addition, we notice that the Alfv\'en velocity tends to be compatible with no reacceleration ($V_A = 0 \units{km/s}$) and that the best-fit halo height is slightly different from that found in the {\it Standard analysis} ($\Delta H \sim 1\units{kpc}$), which means that the antiproton spectrum is also affected by this parameter (and not only to the ratio $D_0/H$).

\begin{figure}[!t]
\centering
\includegraphics[width=0.63\textwidth,height=0.31\textheight,clip]{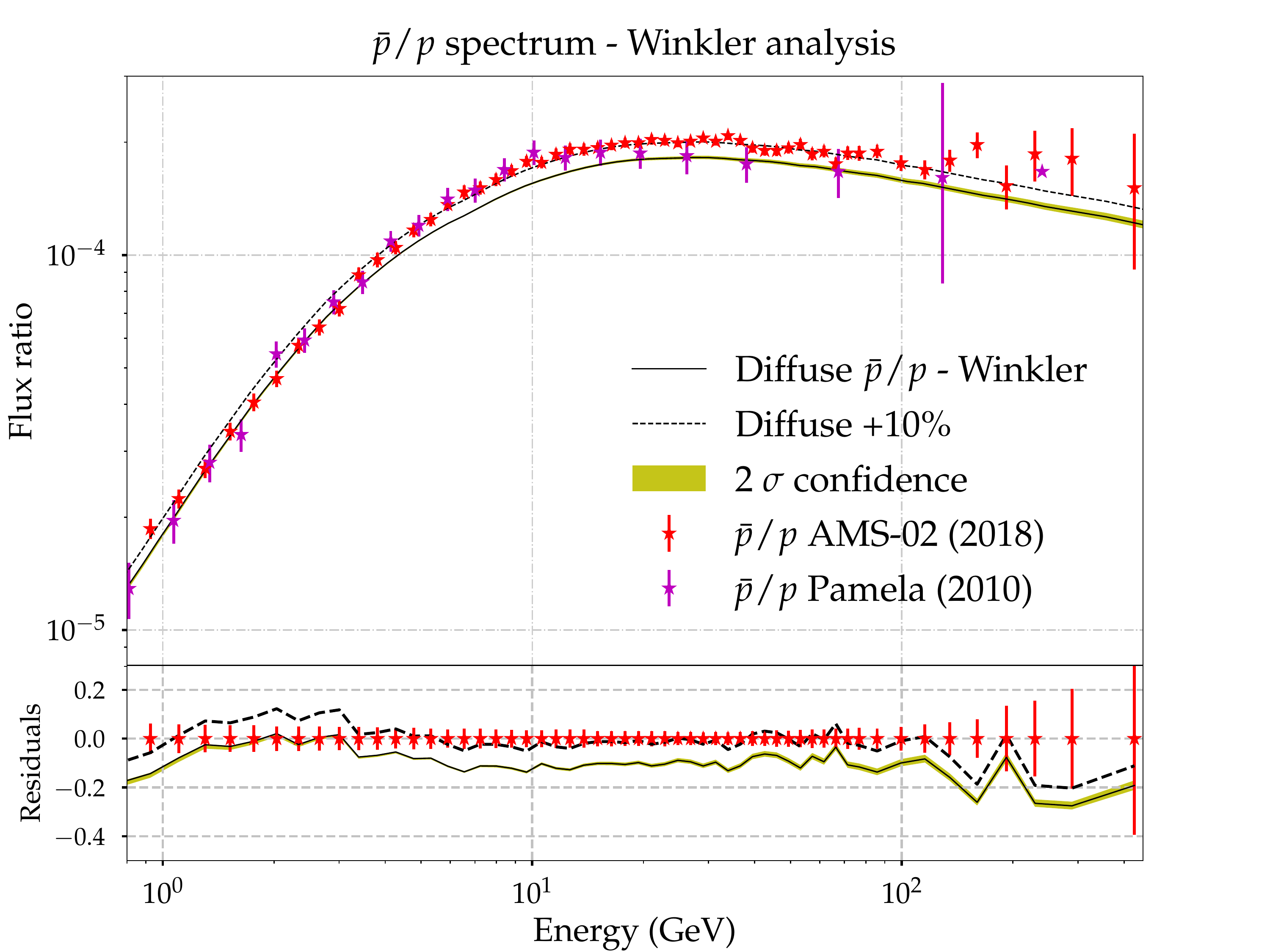}
\caption{$\bar{p}/p$ spectrum evaluated with the propagation parameters determined in the combined $\bar{p}/p$ analysis. The statistical uncertainty in the evaluation of the model is shown as a yellow band and residuals with respect to AMS-02 data (2018) are also shown. The error bars on AMS-02 data are the $1\sigma$ errors reported by the collaboration. On top of this, the same spectrum but scaled by $10\%$ is also shown as a dashed line, allowing us to see that a simple $10\%$ scaling would make us reproduce experimental data within $1\sigma$ errors.}
\label{fig:App_winkler_new}
\end{figure}

Figure~\ref{fig:App_winkler_new} shows the $\bar{p}/p$ spectrum evaluated with these propagation parameters, and using the values $\phi_0 = 0.58\units{GV}$ and $\phi_n = 0.9 \units{GV}$ for the Fisk potential (eq.~\ref{eq:Charge-sign_Modul}). From this evaluation we obtain an important conclusion: the resultant residuals with respect to AMS-02 data are roughly constant and at the level of $\sim10\%$, below the uncertainty associated to the cross sections of production of $\bar{p}$~\cite{Korsmeier}. Therefore, we have also added to this figure a dashed line that represents the same predicted spectrum but scaled by $10\%$ in order to show that it is able to fit the AMS-02 $\bar{p}/p$ experimental data within the reported $1\sigma$ errors above $\sim3\units{GeV}$. We have varied the value of $\phi_n$ from $0.8$ to $1\units{GV}$, finding that the residuals at $10\units{GeV}$ remain the same within $\pm 2\%$. We argue here that, while the fact that the residuals are so low is mainly due to the inclusion of the scaling factors for B, Be and Li production cross sections in the analysis, the new data makes the residuals to be flatter than with the previous $\bar{p}$ AMS-02 data, as shown in more detail in appendix~\ref{sec:old_data}. We argue that the uncertainty related to the contribution of CR species heavier than H and He could also affect this discrepancy by $\sim3\%$.

\subsubsection{Testing possible antiproton production from a WIMP}
\label{sec:Wimptest}

The fact that these residuals are so flat above $3\units{GeV}$ could be related to the effect of cross sections uncertainties, instead of an extra source of antiprotons (such as antiprotons produced from dark matter), whose contribution would have a bump-like structure. In fact, it is very reasonable to think of a $10\%$ rescaling of the cross sections of antiproton production, either from the component of prompt production of antiprotons (direct creation, $CR + gas \rightarrow \bar{p}$) or from the component of antineutron and antihyperons decay ($CR + gas \rightarrow \{\bar{n}, \, \bar{\Lambda}, \, \bar{\Sigma}\} \rightarrow \bar{p}$), or a scaling in both (see, e.g. eq. 5 of ref~\cite{Winkler:2017xor}).
Nevertheless, we also evaluate the spectrum of antiprotons produced from a WIMP, in order to test if this contribution could explain the excess found in the $\bar{p}/p$ spectrum. To do so, we test the mass and thermal annihilation cross sections ($\left< \sigma v \right>$) that yield the best fit with AMS-02 data. This analysis consists of evaluating the spectrum of antiprotons produced from a WIMP annihilating in the $b\bar{b}$ channel and sum this contribution to the diffuse $\bar{p}$ produced from CR collisions (with propagation parameters found in the combined analysis of $\bar{p}/p$ spectrum). The computation of the antiproton spectrum generated from the WIMP is performed with the {\tt DarkSusy} code, for the NFW and Burkert dark matter profiles. Then, we use our MCMC algorithm to find the values of WIMP mass and $\left< \sigma v \right>$ that provide the best fit on the $\bar{p}/p$ ratio.

\begin{figure}[!t]
\centering
\includegraphics[width=0.55\textwidth,height=0.26\textheight,clip] {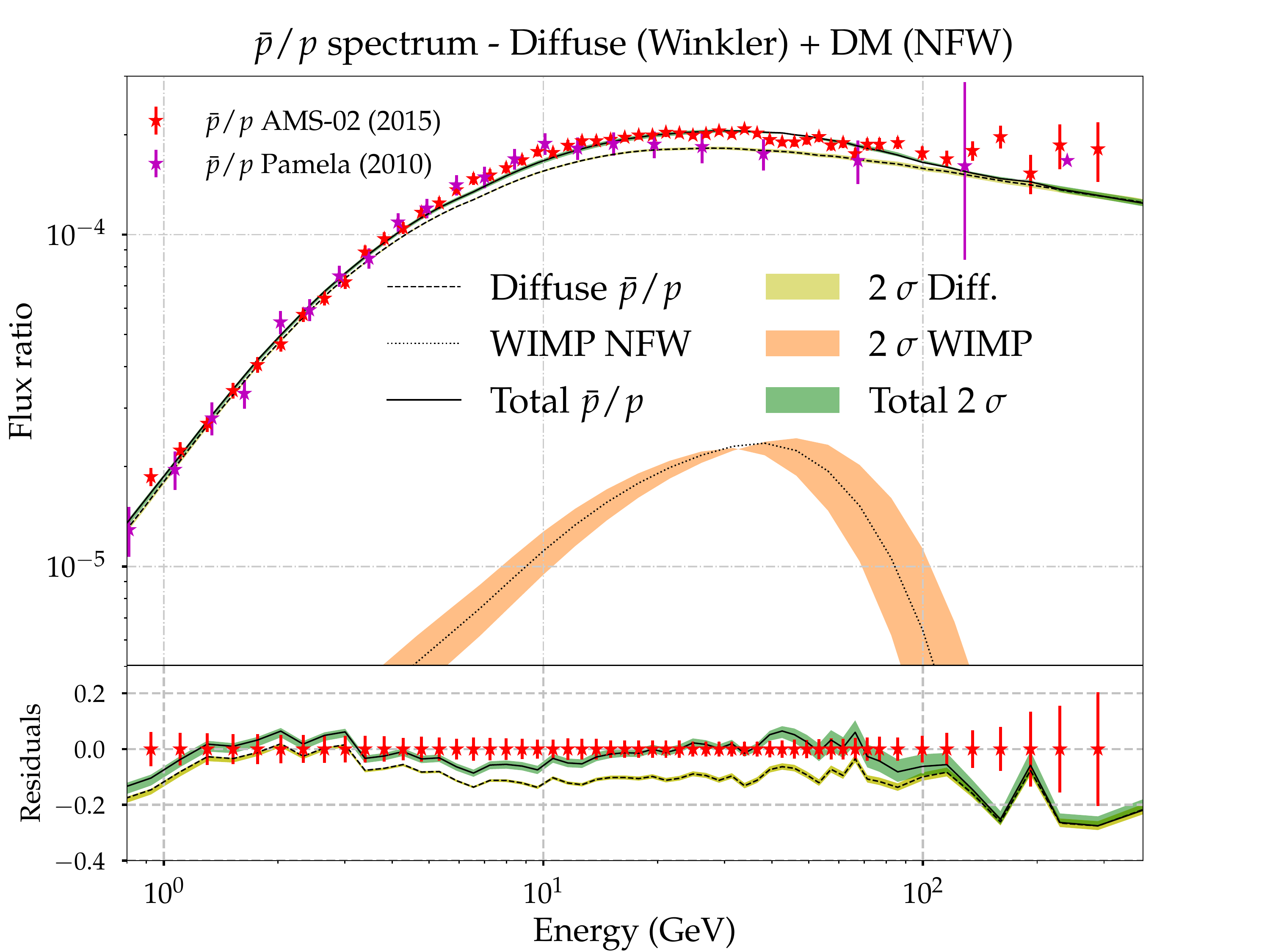}
\includegraphics[width=0.44\textwidth,height=0.245\textheight,clip] {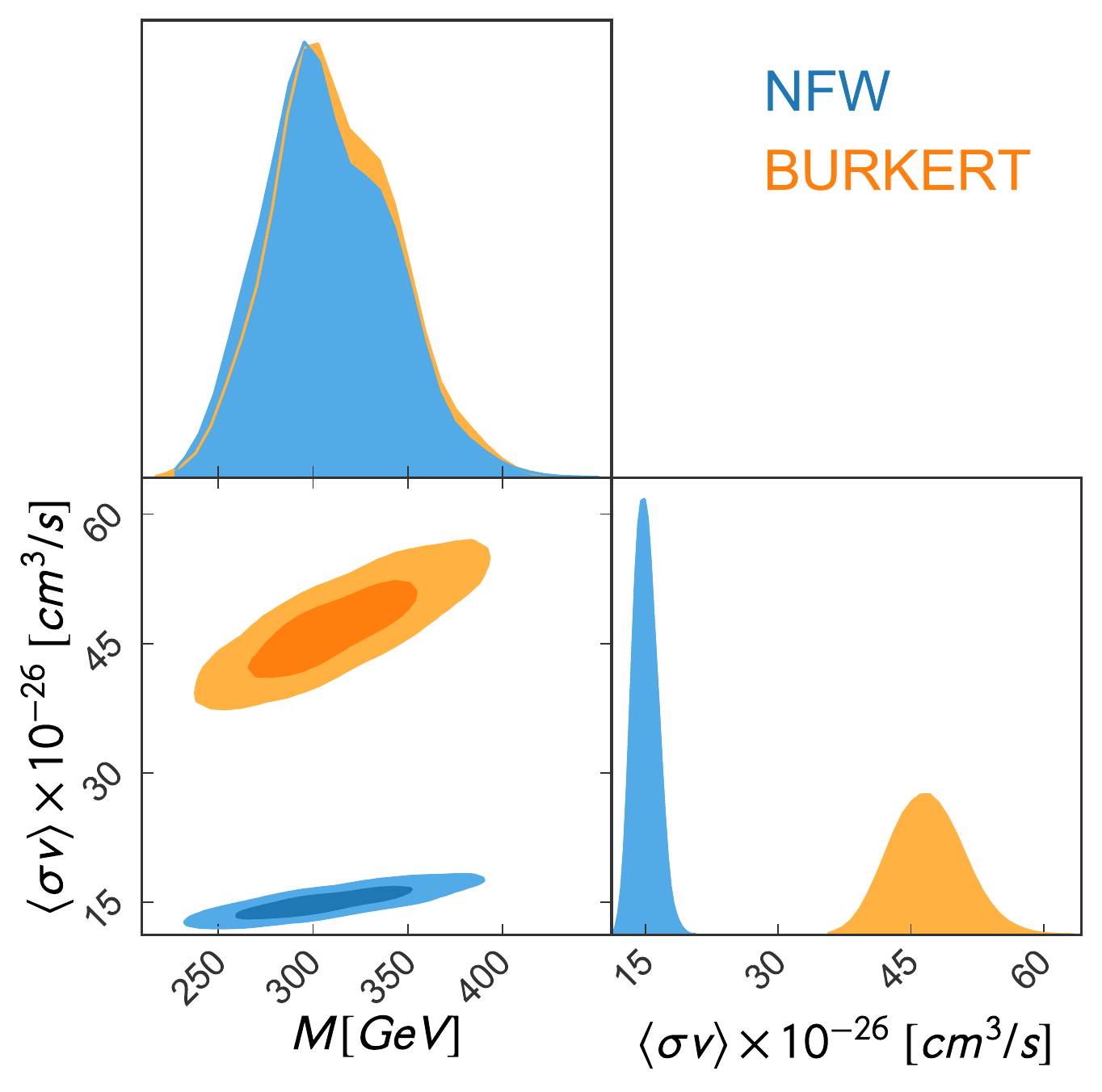}
\caption{Results of the best fit WIMP candidate able to reproduce the discrepancy with respect AMS-02 data. Left panel: the $\bar{p}/p$ spectrum produced from the best WIMP candidate annihilating in the $b\bar{b}$ channel is shown as a dotted line. The uncertainty associated to the determination of mass and $\left< \sigma v \right>$ is shown as an orange band. The diffuse $\bar{p}/p$ spectrum from CR collisions with interstellar gas, found in the combined $\bar{p}/p$ analysis, is shown as a dashed line, with its associated statistical uncertainty as a yellow band. Finally, the sum of both contributions is shown as a solid line and the total statistical uncertainty is shown as a green band.
Right panel: Triangle plot showing the mass and thermally averaged annihilation cross sections of the WIMP candidate that provides the best fit to AMS-02 data. The $68\%$ and $95\%$ uncertainty regions for the Burkert and NFW dark matter density profiles are shown. The best-fit mass value is found to be $304.35^{+8.37}_{-7.31}\units{GeV}$ for the NFW profile and  $ 308.81^{+35.34}_{-27.42}\units{GeV}$ for the Burkert profile. The best-fit $\left<\sigma v \right>$ value is $46.74^{+4.07}_{-3.91}\times10^{-26} \units{cm^{3}/s}$ for the NFW profile and  $ 14.91^{+1.39}_{-1.20}\times10^{-26} \units{cm^{3}/s}$ for the Burkert profile.}
\label{fig:App-Fixed_DM-NFW}
\end{figure}

In the left panel of figure~\ref{fig:App-Fixed_DM-NFW}, we display the result of the analysis of the NFW profile (the WIMP spectrum looks the same for the Burkert profile) in comparison to AMS-02 data. Here, we can see that the component of antiprotons produced from DM annihilation does not seem consistent with the energy dependence of the the $\bar{p}/p$ spectrum, although it can nearly reproduce this spectrum (although antiproton production from two, or more, annihilation channels could fit well with experimental data). We must highlight that the $\chi^2$ value of this fit is $\sim 57$ (evaluated over the $44$ data points of the $\bar{p}/p$ spectrum above $4\units{GeV}$), a higher value than from scaled spectrum, which is $\sim 31$. In the right panel of figure~\ref{fig:App-Fixed_DM-NFW}, the best fit values of WIMP mass and $\left< \sigma v \right>$ are shown as a triangle plot with the $68\%$ and $95\%$ uncertainty regions. The best fit values are of $\sim300 \units{GeV}$ for both profiles and a value of the thermal annihilation cross sections of $\sim 1.5 \times 10^{-25} \units{cm^{3}/s}$ for the NFW profile and $\sim 4.7 \times 10^{-25} \units{cm^{3}/s}$ for the Burkert profile. These values are much larger than the values usually quoted in indirect searches of DM from antiprotons (as seen in appendix~\ref{sec:old_data} this is mainly related to the use of the new $\bar{p}$ AMS-02 data) and well above with calculations of the thermal relic cross sections~\cite{Steigman:2012nb}. In addition, these values are in tension or above the Fermi-LAT constraints of Dwarf galaxies~\cite{DM_limits} and Milky Way satellites~\cite{Fermi-LAT:2016uux}, and are not compatible with the possible dark matter signal from the Galactic Center Excess (GCE)~\cite{DiMauro:2021raz}. However, these values are neither discarded from Planck CMB constraints nor from AMS-02 constraints on positrons, as shown in ref~\cite{DM_limits}. 
It is also important to mention that, even though other dark matter annihilation channels would change the mass and $\left< \sigma v \right>$ predicted, the other hadronic channels ($t\bar{t}$, $c\bar{c}$ and $s\bar{s}$) produce a very similar flux of antiprotons at $10 \units{GeV}$ and the $W^{+}W^{-}$ channel leads to a slightly larger $\left< \sigma v \right>$ to reproduce the $\bar{p}/p$ excess.



The direct inputs and outputs from the DRAGON2 runs with the propagation parameters obtained from both analyses are available at this \href{https://github.com/tospines/Analyses-and-plotting-codes/tree/main/DRAGON2_Antiprotons} {repository}\footnote{\url{https://github.com/tospines/Analyses-and-plotting-codes/tree/main/DRAGON2_Antiprotons}}, as well as the best-fit spectra of proton, helium and antiproton. Moreover, we include in the repository the tables of antiproton flux produced from the annihilation of WIMPs for diverse masses and $\left< \sigma v \right>$ values in the $b\bar{b}$ channel for the two DM density distributions studied in this work and show a comparison between the flux of antiprotons produced from annihilation from all the hadronic channels and the $W^{+}W^{-}$ channel at the best-fit mass and $\left< \sigma v \right>$ found for the NFW distribution.

\section{Discussion and conclusions}
\label{sec:conc}

In this work, we have studied the diffuse $\bar{p}/p$ spectrum produced in CR interactions with the interstellar gas. We first carry out a combined analysis, without including the $\bar{p}/p$ ratio, in which we have taken into account different secondary CRs and their ratios, as well as nuisance parameters that allow us to renormalize the cross sections of production of B, Be and Li. These observables allow us to determine the propagation parameters (H, $D_0$, $V_A$, $\eta$ and $\delta$) to provide a simultaneous fit of the ratios of the different CR species. We show that the $\bar{p}/p$ spectrum predicted from this analysis differ from the AMS-02 data by a factor of $20-30\%$. 

Then, we include the $\bar{p}/p$ spectrum in our combined analysis, making use of the new AMS-02 data for this spectrum. This analysis yield a much closer prediction of the $\bar{p}/p$ spectrum with respect to recent AMS-02 data, producing discrepancies that are constant above $3\units{GeV}$ and are of $\sim10\%$. Interestingly, the new AMS-02 data lead to flatter residuals than with the previous dataset. We discuss that this discrepancy seems to be plausibly explained by just a $10\%$ scaling of the cross sections of $\bar{p}$ production, which makes our model match completely the AMS-02 $\bar{p}/p$ experimental data within $1\sigma$ uncertainties. To test alternative explanations of this excess, we evaluate if an extra component of antiprotons from annihilation of a WIMP could reproduce this discrepancy. We find that, although this component can be close to reproduce the $\bar{p}/p$ spectrum, the values of mass and thermally averaged annihilation cross sections (the best fit values found are: WIMP mass $\sim300\units{GeV}$ and $\left<\sigma v \right> \sim 1.5-4.7\times10^{-25} \units{cm^{3}/s}$) are in tension or ruled out by several other indirect dark matter searches. Also the $\chi^2$ value obtained from the fit with a $10\%$ scaling is almost a factor of two lower than the fit considering DM annihilating in antiprotons.

Therefore, we conclude that taking into account all the sources of uncertainties in the evaluation of the secondary antiprotons produced from CR interactions with gas allows as to explain the $\bar{p}/p$ spectrum without any need of an extra source of production of $\bar{p}$. Specifically, it seems that the energy dependence of the $\bar{p}/p$ spectrum is well reproduced assuming a pure secondary origin of antiprotons and that the excess found is plausibly explained by a rescaling of the cross sections of $\bar{p}$ production. On the contrary, it seems difficult that the component of WIMP annihilation alone could be the responsible of this excess, although this source of antiprotons could be still present, below the high level of uncertainties associated to the evaluation of the $\bar{p}$ spectrum.

In this way, we provide a set of propagation parameters and scale factors for renormalizing the cross sections parametrizations that allow us to reproduce all the ratios of B, Be, Li and $\bar{p}$ simultaneously.
 
Finally, we highlight the most important points revealed of this work: 
\begin{itemize}
    \item An analysis that takes into account uncertainties in the production cross sections of the main secondary CRs B, Be and Li leads to residuals in the $\bar{p}/p$ spectrum always lower than $12\%$ at $10\units{GeV}$. The total uncertainties are, at least, a factor of two greater.

    \item The use of the new $\bar{p}/p$ AMS-02 data has an important impact in the conclusions reached, since the residuals found are much flatter than when the old dataset is employed. 
    
    \item  The shape of the reported excess does not resemble a bump, as would be expected from  dark matter annihilation. In addition, the best-fit WIMP candidate obtained with this new dataset seems to be incompatible with most of the indirect searches of dark matter currently done. The fact that we obtain a much larger WIMP mass than the one predicted in previous works is mainly due to the use of the new AMS-02 $\bar{p}/p$ data.
\end{itemize}

\subsection{Comparison with other works}
\label{sec:Works}
Many recent works have been dedicated to the study of the antiprotons produced from CR collisions and the possible signature of antiprotons produced from DM. The main difference with respect to previous analyses is the use of the scale factors for renormalizing the cross sections of production of the main secondary CRs and the use of the new AMS-02 data. 

Previous works have found an excess of antiprotons with maximum residuals around $10\units{GeV}$, with a bump-like structure similar to what we find in the analysis with the old AMS-02 data (fig.~\ref{fig:App_Winkler_old}). Examples of these studies are refs.~\cite{Cuoco, Cholis, Cuoco2}. However, they did not take into account the full systematic uncertainties in the evaluation of the $\bar{p}$ production due to the uncertainties in spallation cross sections of B (since they use only the B/C ratio). In addition, they predict a WIMP mass close to $100\units{GeV}$ to explain the antiproton excess, which would have changed significantly (towards larger WIMP mass) using the new AMS-02 dataset.

Some other works claim that this excess is not significant at all under the systematic uncertainties. This is the case of, for example, refs.~\cite{Winkler:2017xor, Boudaud_PRR, Heisig}, where the authors claim that the $\bar{p}$ spectrum is compatible with a pure secondary origin when taking into account all the systematic sources of uncertainty. The two last references also include possible correlations in the AMS-02 data from a guessed correlation matrix. This is something we do not include in this study, since we employ only the errors reported by the AMS-02 collaboration, and would have the effect of broadening the uncertainties associated to the determination of the propagation parameters. 

\acknowledgments

We are very grateful to the DAMASCO group, at the instituto de física teórica (IFT) in Madrid, for their support and valuable conversations related to this work. We thank a lot Tim Linden and Daniele Gaggero for the valuable comments on the manuscript.
Thanks to Torsten Bringmann for his help in some technical aspects on the DarkSusy code.
This work has been carried out using the RECAS computing infrastructure in Bari (\url{https://www.recas-bari.it/index.php/en/}). A particular acknowledgment goes to G. Donvito and A. Italiano for their valuable support. Pedro de la Torre Luque is supported by the Swedish Research Council under contract 2019-05135.

\newpage

\bibliographystyle{apsrev4-1}
\bibliography{biblio}

\newpage
\appendix
\section{Results with the previous AMS-02 $\bar{p}/p$ dataset}
\label{sec:old_data}

In this appendix, we show the results of the combined analysis including the old AMS-02 data for the $\bar{p}/p$ ratio (data collected in the period 2011-2015). We apply the same analysis as in section~\ref{sec:Ap_analysis}, with the difference that the Fisk potential employed here is set to $\phi_0 = 0.61 \units{GV}$ also for the $\bar{p}/p$ spectrum (corresponding to the period of data collection of the previous $\bar{p}/p$ dataset).

\begin{figure}[!th]
\centering
\includegraphics[width=0.49\textwidth,height=0.25\textheight,clip] {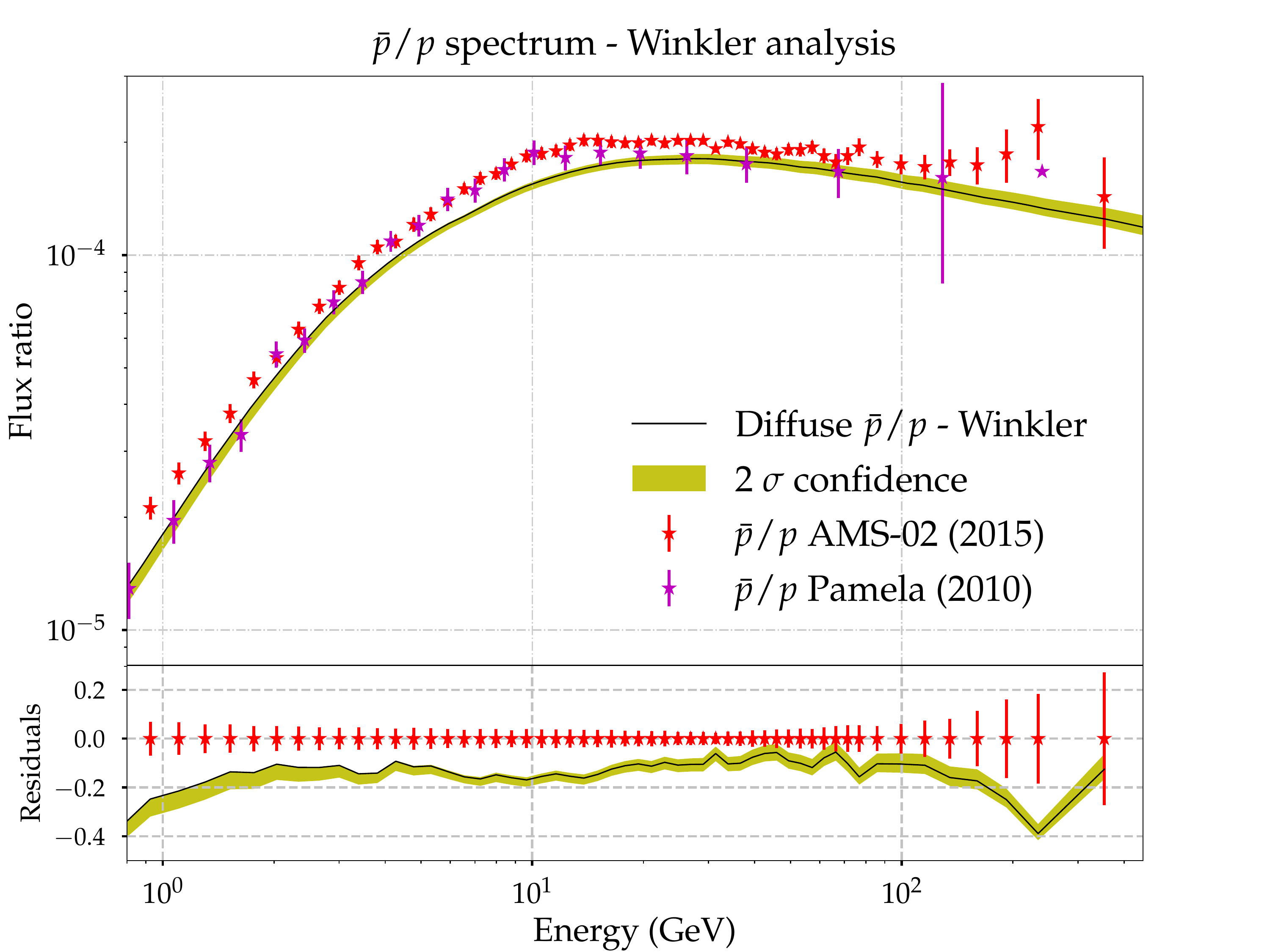} 
\hspace{-0.65 cm} \includegraphics[width=0.49\textwidth,height=0.25\textheight,clip] {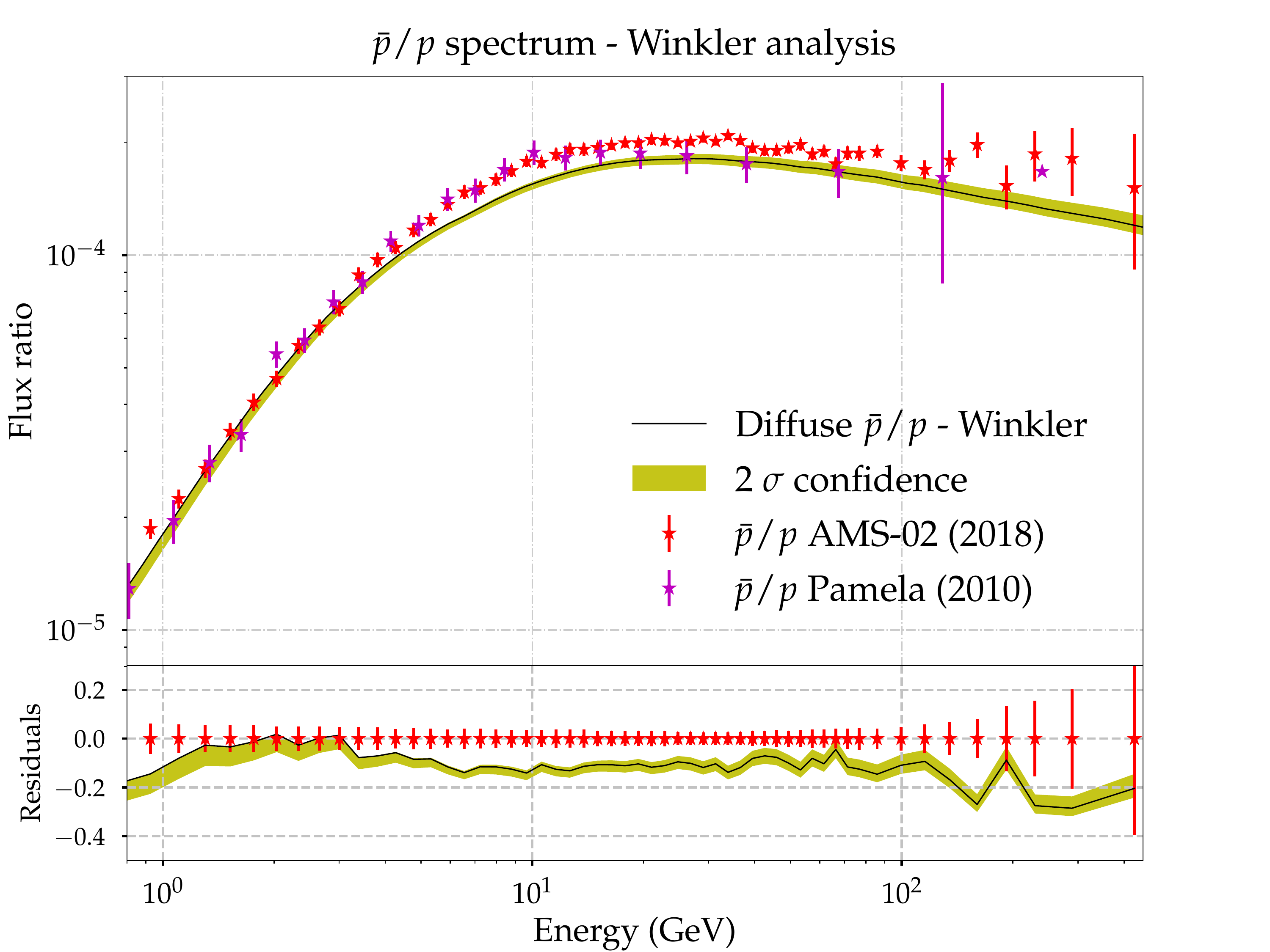} 
\caption{Left panel: $\bar{p}/p$ spectrum evaluated with the propagation parameters determined in the combined $\bar{p}/p$ analysis with the old AMS-02 dataset (2011-2015). The statistical uncertainty in the evaluation of the model is shown as a yellow band and residuals with respect to AMS-02 data (2015) are also shown. Right panel: same as in the left panel but using the recently published dataset (2011-2018). The evaluated $\bar{p}/p$ spectrum is the same in both cases.}
\label{fig:App_Winkler_old}
\end{figure}

Relevantly, the propagation parameters found in the analysis with the old $\bar{p}/p$ dataset are roughly identical to those obtained using the new AMS-02 dataset (see table~\ref{tab:diff_params}), as expected. In figure~\ref{fig:App_Winkler_old}, we show the predicted spectrum evaluated with the Winkler cross sections and the propagation parameters labeled as ``$\bar{p}/p$ old data'' in table~\ref{tab:diff_params}. In the leftpanel, the experimental data are those from the previous AMS-02 dataset (data collection from 2011-2015), while in panel b they are the newly published data (data collection from 2011-2018). A clear change in the shape of the residuals is obtained, even though the evaluated spectrum is the same in both cases. The new dataset leads to a flattening of the shape of the residuals and the WIMP signal which is able to explain this discrepancy is shifted towards larger masses. 

\begin{figure}[!th]
\centering
\includegraphics[width=0.55\textwidth,height=0.26\textheight,clip] {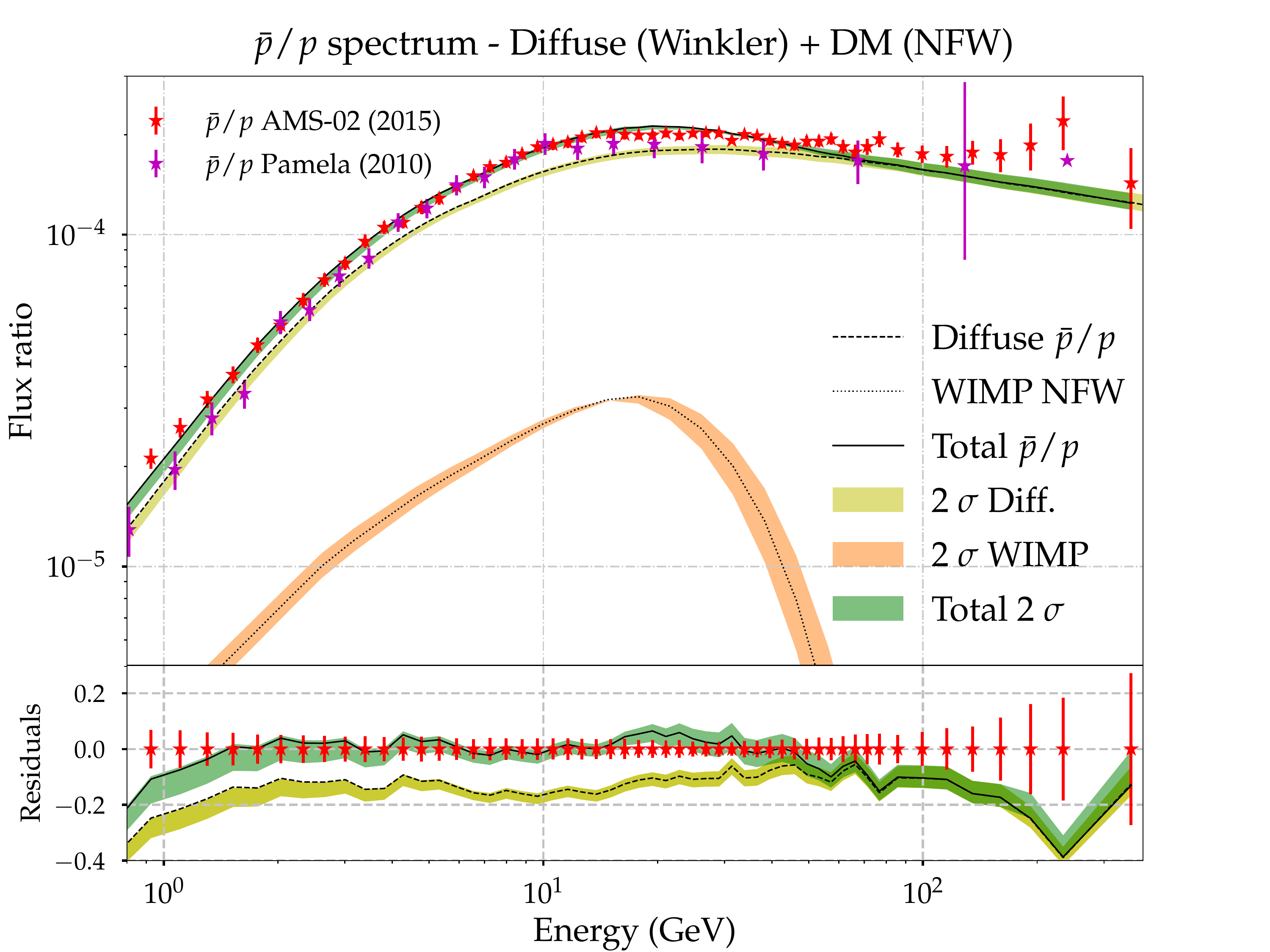}
\includegraphics[width=0.44\textwidth,height=0.245\textheight,clip] {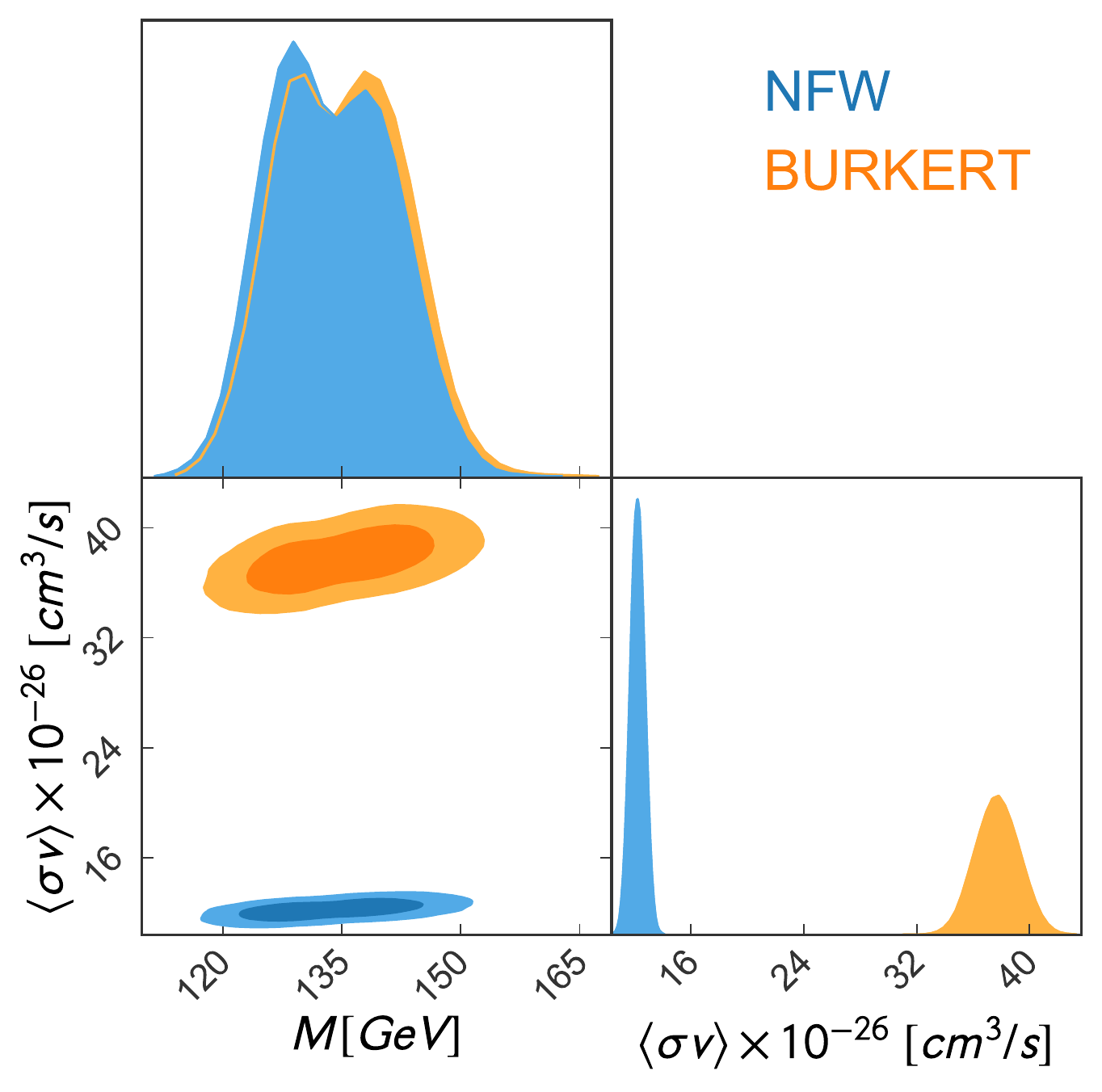}
\caption{Same as in figure~\ref{fig:App-Fixed_DM-NFW}, but for the WIMP candidate found in the analysis with the previous $\bar{p}/p$ AMS-02 dataset. The best-fit mass value is found to be $ 133.3^{+8.37}_{-7.31}\units{GeV}$ for the NFW profile and  $ 135.16^{+7.73}_{-8.02}\units{GeV}$ for the Burkert profile. The best-fit $\left<\sigma v \right>$ value is $12.21^{+0.49}_{-0.48}\times10^{-26} \units{cm^{3}/s}$ for the NFW profile and  $ 37.72^{+1.52}_{-1.51}\times10^{-26} \units{cm^{3}/s}$ for the Burkert profile.}
\label{fig:App-Fixed_DM-NFW_old}
\end{figure}

In figure~\ref{fig:App-Fixed_DM-NFW_old}, we repeat the exercise shown in figure~\ref{fig:App-Fixed_DM-NFW}, but for the WIMP signal that achieves the best fit the old $\bar{p}/p$ dataset. Panel a shows the evaluated spectrum with the contribution of a WIMP producing antiprotons (WIMP annihilating into the $b\bar{b}$ channel) which provides the best fit to this experimental data. In the right panel, we show the $1$ and $2\sigma$ significance regions of the probability distribution function of the inferred mass and $\left<\sigma v \right>$, for the Burkert and NFW dark matter profiles in the Galaxy. While the inferred $\left<\sigma v \right>$ value (equal to $\sim1.2\units{cm^{3}/s}$ for the NFW profile and $\sim3.7\units{cm^{3}/s}$ for the Burkert profile) is slightly lower than the one obtained in the analysis with the new dataset, the inferred mass is significantly lower (from $\sim 300\units{GeV}$ to $\sim 130\units{GeV}$).
In fact, the WIMP signal necessary to reproduce the $\bar{p}/p$ spectrum with the new experimental data is much less credible than with the previous dataset.

\FloatBarrier
\textfloatsep 2.5cm
\section{Summary of the MCMC results: propagation parameters}
\label{sec:diff_params}
\begin{figure}
\centering
\includegraphics[width=0.95\textwidth,height=0.6\textheight,clip]{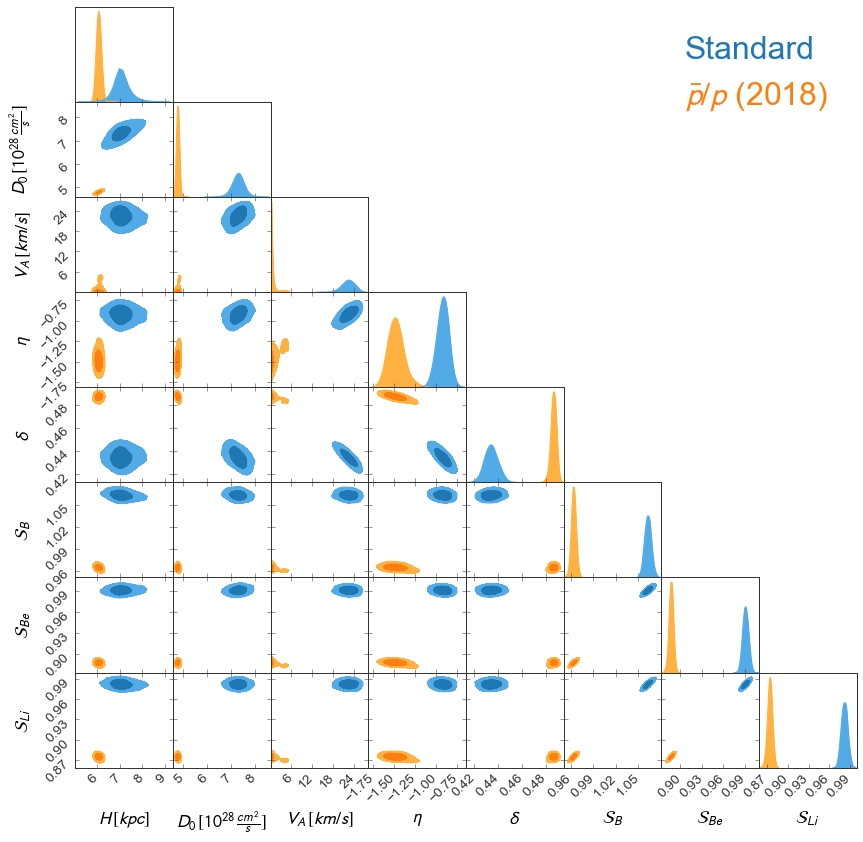} 
\caption{Probability distributions of the considered propagation parameters obtained for the main analyses performed in this work. The contour plots highlight the $68\%$ and $95\%$ credible regions.}
\label{fig:joint_PDFs}
\end{figure}

In this appendix, we report the diffusion parameters obtained for the standard and combined $\bar{p}/p$ analyses in table~\ref{tab:diff_params}, which contains the median values (maximum posterior probability value) $\pm$ the $1\sigma$ uncertainty related to their determination and the actual range of values contained in the 95\% probability of the distribution. We also show the probability distribution functions (PDFs) of each propagation parameter obtained from the different analyses in figure~\ref{fig:joint_PDFs}.

\vspace{0.7cm}
\begin{table}[htb!]
\centering
\resizebox*{0.9\columnwidth}{0.17\textheight}{
\begin{tabular}{|c|c|c|c|c|c|c|c|c|}
\hline
\multicolumn{9}{|c|}{\textbf{\large{Propagation parameters and scale factors}}} \\
\hline &\textbf{$H$ (kpc}) & \textbf{$D_0$ \, ($10^{28}$ cm$^{2}$/s)} & \textbf{$v_A$ (km/s)} & \textbf{$\eta$} & \textbf{$\delta$} & \textbf{$\mathcal{S}_{B}$} & \textbf{$\mathcal{S}_{Be}$} & \textbf{$\mathcal{S}_{Li}$} \\ 
\hline
\multirow{2}{7em}{\centering Standard} & 7.05 $\pm$ 0.3 & 7.32$\pm$ 0.22   &  22.47 $\pm$ 1.81 &  -0.92 $\pm$ 0.07 & 0.43  $\pm$ 0.01 &  1.06 $\pm$ 0.01 &  0.99 $\pm$ 0.01 &  0.98 $\pm$ 0.01 \\  &  [7.71, 6.57]  &   [7.75, 6.90] &   [25.97, 18.73]  &  [-1.16, -0.77] &   [0.41, 0.45] &   [1.04, 1.08] &   [0.97, 1.01] & [0.96, 1.]\\ \hline 

\multirow{2}{6em}{\centering $\bar{p}/p$ new data (2018)}  & 6.07 $\pm$ 0.11 & 4.79$\pm$ 0.1   &  0.28 $\pm$ 1.25 &  -1.57 $\pm$ 0.08 & 0.49  $\pm$ 0.01 &  0.96 $\pm$ 0.01 &  0.89 $\pm$ 0.01 &  0.87 $\pm$ 0.01 \\  &  [5.82, 6.27]  &   [4.59, 5.01] &   [0., 2.8]  &  [-1.75, -1.39] &   [0.46, 0.51] &   [0.94, 0.99] &   [0.87, 0.91] & [0.85, 0.90]\\ \hline 

\multirow{2}{6em}{\centering $\bar{p}/p$ old data (2015)}  & 6.04 $\pm$ 0.11 & 4.79$\pm$ 0.06   &  0.28 $\pm$ 1.25 &  -1.57 $\pm$ 0.09 & 0.49  $\pm$ 0.01 &  0.97 $\pm$ 0.01 &  0.89 $\pm$ 0.01 &  0.89 $\pm$ 0.01 \\  &  [5.82, 6.27]  &   [4.59, 5.01] &   [0., 2.8]  &  [-1.75, -1.39] &   [0.46, 0.51] &   [0.95, 0.99] &   [0.87, 0.91] & [0.87, 0.91]\\ \hline 
\end{tabular}
}

\caption{Summary of the results obtained in the different analyses performed in this work. The error given corresponds to the $1\sigma$ uncertainty assuming the PDF to follow a Gaussian distribution. The actual $2\sigma$ uncertainty range is given for every parameter within square brackets.}
\label{tab:diff_params}
\end{table}

\end{document}